\documentclass[useAMS,usenatbib]{mn2e}

\usepackage{graphicx}
\usepackage{multirow}

\usepackage{array}
\newcolumntype{$}{>{\global\let\currentrowstyle\relax}}
\newcolumntype{^}{>{\currentrowstyle}}
\newcommand{\rowstyle}[1]{\gdef\currentrowstyle{#1}%
  #1\ignorespaces
}

\title[Star formation in A2744]{Star formation in the massive cluster merger Abell 2744}
\author[T.~D.~Rawle et al.]
{\parbox{\textwidth}{
T.~D.~Rawle,$^{1}$\thanks{E-mail: \texttt{tim.rawle@sciops.esa.int}}
B.~Altieri,$^{1}$
E.~Egami,$^{2}$
P.~G.~P\'{e}rez-Gonz\'{a}lez,$^{3}$
J.~Richard,$^{4}$
J.~S.~Santos,$^{5,1}$
I.~Valtchanov,$^{1}$
G.~Walth,$^{2}$
H.~Bouy,$^{6}$
C.~P.~Haines,$^{7}$
N.~Okabe$^{8}$
}\vspace{0.4cm}\\
\parbox{\textwidth}{
$^{1}$ESAC, ESA, PO Box 78, Villanueva de la Ca\~{n}ada, 28691 Madrid, Spain\\
$^{2}$Steward Observatory, University of Arizona, 933 N. Cherry Ave, Tucson, AZ 85721, USA\\
$^{3}$Departamento de Astrof\'isica, Facultad de CC. F\'isicas,Universidad Complutense de Madrid, E-28040 Madrid, Spain\\
$^{4}$CRAL, Observatoire de Lyon, Universit\'e de Lyon 1, 9 avenue Ch. Andr\'e, F-69561 Saint-Genis Laval, France\\
$^{5}$Osservatorio Astrofisico di Arcetri, Largo Enrico Fermi 5, I-50125 Firenze, Italy\\
$^{6}$Centro de Astrobiolog\'ia, INTA-CSIC, PO Box 78, Villanueva de la Ca\~{n}ada, 28691 Madrid, Spain\\
$^{7}$Departamento de Astronom\'ia, Universidad de Chile, Casilla 36-D, Correo Central, Santiago, Chile\\
$^{8}$Kavli Institute for the Physics and Mathematics of the Universe (WPI), Todai Institutes for Advanced Study, University of Tokyo, 5-1-5 Kashiwanoha, Kashiwa, Chiba 277-8583, Japan\\
}}
\begin{document}

\date{28 February 2014}

\pagerange{\pageref{firstpage}--\pageref{lastpage}} \pubyear{2013}

\maketitle

\label{firstpage}

\begin{abstract}
We present a comprehensive study of star-forming (SF) galaxies in the \textit{HST Frontier Field} recent cluster merger A2744 ($z=0.308$). Wide-field, ultraviolet--infrared (UV--IR) imaging enables a direct constraint of the total star formation rate (SFR) for 53 cluster galaxies, with SFR$_{\rm UV+IR}$ $=$ $343\pm10$~$M_{\sun}$~yr$^{-1}$. Within the central 4~arcmin (1.1~Mpc) radius, the integrated SFR is complete, yielding a total SFR$_{\rm UV+IR}$ $=$ $201\pm9$~$M_{\sun}$~yr$^{-1}$. Focussing on obscured star formation, this core region exhibits a total SFR$_{\rm IR}$ $=$ $138\pm8$~$M_{\sun}$~yr$^{-1}$, a mass-normalised SFR$_{\rm IR}$ of $\Sigma_{\rm SFR} = 11.2\pm0.7$~$M_{\sun}$~yr$^{-1}$ per 10$^{14}$~$M_{\sun}$ and a fraction of IR-detected SF galaxies $f_{\rm SF} = 0.080^{+0.010}_{-0.037}$. Overall, the cluster population at $z\sim0.3$ exhibits significant intrinsic scatter in IR properties (total SFR$_{\rm IR}$, $T_{\rm dust}$ distribution) apparently unrelated to the dynamical state: A2744 is noticeably different to the merging Bullet cluster, but similar to several relaxed clusters. However, in A2744 we identify a trail of SF sources including jellyfish galaxies with substantial unobscured SF due to extreme stripping (SFR$_{\rm UV}$/SFR$_{\rm IR}$ up to 3.3). The orientation of the trail, and of material stripped from constituent galaxies, indicates that the passing shock front of the cluster merger was the trigger. Constraints on star formation from both IR \textit{and} UV are crucial for understanding galaxy evolution within the densest environments.
\end{abstract}

\begin{keywords}
galaxies: clusters: individual: Abell 2744 -- galaxies: star formation -- infrared: galaxies
\end{keywords}

\section{Introduction}
\label{sec:intro}

Galaxy clusters evolve through secular growth, via accretion of galaxies (or small groups) along filaments, interspersed by occasional violent mergers. When first encountering the dense intracluster medium (ICM), an in-falling galaxy may experience triggered star formation \citep{bek99-15,koy08-1758}, but starvation and ram-pressure stripping soon (within the first pass; \citealt{tre03-53}) prevent further activity \citep{gun72-1,bos06-517,hai13-126,raw13-2667}. Detailed analysis of individual nearby cluster galaxies reveals simultaneous starbursts and quenching \citep{mer13-1747}, while UV observations confirm that asymmetric ``jellyfish'' morphologies, long tails and knots of star formation, result from gas stripping during cluster in-fall, and conclude that the phenomenon is widespread in massive clusters \citep{smi10-1417}. At higher redshift, only a handful of jellyfish galaxies are known (e.g. \citealt{owe12-23}, \citealt{ebe14-40}) due to the difficulty in identifying often small, faint features.

In relaxed clusters, galaxies in the core are generally quiescent (although c.f. brightest cluster galaxies, \citealt{raw12-29}), and star formation is concentrated in the periphery \citep{fad08-9}. The core-passage phase of a massive cluster merger may produce a shock front moving through the ICM that could initiate further starbursts in central galaxies \citep[e.g.][]{bek03-13} or curtail star formation via extreme ram-pressure stripping \citep[e.g.][]{fuj99-1}. The relative significance of these processes is not currently well constrained \citep{owe05-31,joh08-289,raw12-106,owe12-23}.

An understanding of the influence of cluster mergers requires a comprehensive view of star formation in cluster galaxies. The local Universe lacks recent massive mergers for a detailed analysis. The Bullet cluster is a famous example at intermediate redshift ($z\sim0.3$), presenting a massive supersonic merger in the plane of the sky. IR analysis shows a marginal increase in total SFR compared to relaxed clusters at similar redshift, but no obvious individual enhancement in the vicinity of the shock front \citep{chu10-1536,raw10-14,raw12-106}. However, IR data only probes obscured star formation, which may miss activity in galaxies undergoing gas (and therefore dust; \citealt{cor10-49}) stripping. Unfortunately, wide-field UV imaging (tracing unobscured star formation) is non-existent for the Bullet cluster due in part to a \textit{GALEX} bright star constraint.

In this paper we explore star formation in Abell 2744 (00$^{\rm h}$14$^{\rm m}$19$^{\rm s}$, --30$^{\circ}$23'19'', $z=0.308$), also known as AC118 or  ``Pandora's Cluster''. A2744 includes a recent supersonic merger approximately in the plane of the sky, thus offering an alternative laboratory to probe the effect of cluster-scale shocks on galaxy formation. Analysis of A2744 benefits from UV--IR coverage and many 100s of cluster galaxy spectroscopic redshifts, which led to the cluster's selection as the first \textit{HST Frontier Field} (HFF).\footnote{http://www.stsci.edu/hst/campaigns/frontier-fields/}

Early optical studies agreed with the predicted blue galaxy fraction from the \citet{but84-426} effect \citep{cou87-423}. The merger scenario was not proposed until discovery of a radio relic \citep{gio99-141} and a non-gaussian velocity distribution \citep{gir01-79}. \textit{Chandra} X-ray imaging then helped identify two remnant cores from a recent, $\sim$3:1 mass ratio, `bullet'-like merger \citep{kem04-385,bos06-461,bra09-947,owe11-27,mer11-333}. Debris stripped from the more massive northern core is found between the remnants, while the outskirts comprise the mixed, de-coupled halos of the pre-merger systems. A third component lies to the northwest, originally characterised as a pre-infall sub-cluster \citep{kem04-385,bos06-461}, but recently re-interpreted as a post-stripped system moving north-northeast \citep{owe11-27} or northwest \citep{mer11-333}. Two large-scale filaments extend to the south and northwest \citep{bra09-947}.

We explore the cluster galaxy (UV+IR) SFRs in the context of this substructure. Section \ref{sec:obs} overviews the observations and Section \ref{sec:analysis} derives SFRs. Section \ref{sec:results} discusses star formation properties of individual galaxy and the integrated cluster. Section \ref{sec:conclusions} summarises the results. We adopt standard cosmological parameters ($\Omega_{\rm M}$,$\Omega_{\Lambda}$,$h$) = (0.3,0.7,0.7). At the cluster redshift, 1~arcmin corresponds to 0.27~Mpc.

\begin{figure*}
\centering
\includegraphics[width=177mm]{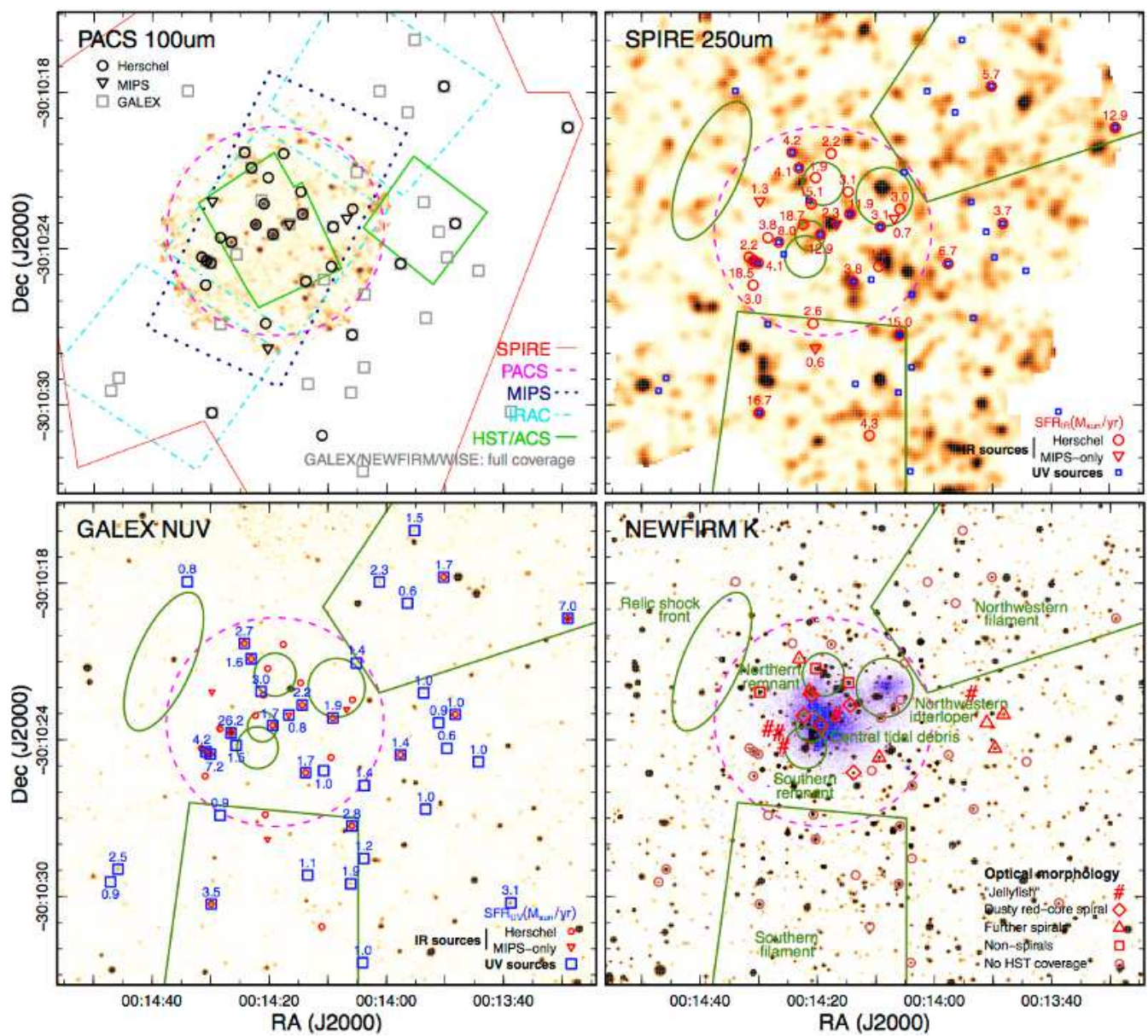}
\caption{A2744 cluster field, highlighting star-forming spectroscopic cluster members ($0.273 \le z \le 0.339$). The dashed magenta circle indicates a 4~arcmin (1.1~Mpc) radius. In three panels, cluster substructure is indicated in green (labelled in the lower--right panel). \textit{Upper--left:} PACS 100~$\mu$m map over-plotted by all data footprints. \textit{Upper--right:} SPIRE 250~$\mu$m map with IR-detected members labelled by SFR$_{\rm IR}$ ($M_{\sun}$~yr$^{-1}$). \textit{Lower--left:} \textit{GALEX} $NUV$ image (smoothed to 5~arcsec FWHM), with UV-detected members labelled by (uncorrected) SFR$_{\rm UV}$ ($M_{\sun}$~yr$^{-1}$). \textit{Lower--right:} NEWFIRM $K_{\rm s}$-band overlaid by the \textit{Chandra} X-ray map (blue; 2~arcsec Gaussian smoothing). Star-forming cluster galaxies are classified by optical morphology (Section \ref{sec:morphology}).}
\label{fig:field}
\end{figure*}

\section{Observations}
\label{sec:obs}

This comprehensive study of star formation requires spectroscopic redshifts, well-sampled infrared SEDs (indicating dust-obscured star formation) and photometry in UV bands (tracing unobscured young stellar populations).

\subsection{\textit{Herschel} observations}

A2744 was targeted by the \textit{Herschel Space Observatory} \citep{pil10-1} as part of the ``\textit{Herschel} Lensing Survey'' \citep[HLS;][]{ega10-12}. PACS images at 100 and 160~$\mu$m \citep{pog10-2} were produced from calibrated time-stream data, taken directly from the \textit{Herschel} Science Archive.\footnote{http://herschel.esac.esa.int/Science\_Archive.shtml} Map-making followed the Generalised Least Squares method of Unimap \citep{pia12-3687}. For SPIRE \citep[250, 350, 500~$\mu$m;][]{gri10-3}, images were created via the standard reduction pipeline in HIPE v9, plus a median baseline removal and the \textit{fastDestriper} module.

All maps are centred on the cluster core, extending to radii of $\sim$4 and 9~arcmin for PACS and SPIRE respectively. Figure \ref{fig:field} (upper row) displays the 100 and 250~$\mu$m images, smoothed by the relevant beam size (7.7, 12, 18, 25, 36~arcsec FWHM for 100--500~$\mu$m). Mean 5-$\sigma$ sensitivities are 4.4, 8.7, 11, 13 and 12~mJy, although the three SPIRE bands are confusion limited: 5$\,\sigma_{\rm conf}$ $\approx$ 28, 32, 33~mJy \citep{ngu10-5}. Further details are presented in HLS survey papers.

\subsection{Additional infrared imaging}

\textit{Spitzer} data were extracted from the Heritage Archive.\footnote{http://irsa.ipac.caltech.edu/applications/Spitzer/SHA} IRAC four-band imaging (3.6, 4.5, 5.8, 8~$\mu$m) consists of three $6\times6$~arcmin regions: one central and two flanking. MIPS 24~$\mu$m data ($6\times11$~arcmin) covers the entire central IRAC region. \textit{Spitzer} beam sizes (increasing wavelength) are 1.7, 1.7 1.7, 1.9, 6.0~arcsec FWHM, and mean 5-$\sigma$ sensitivities are 1.4, 1.5, 8.2, 7.2, 89~$\mu$Jy.

The all-sky \textit{WISE} \citep{wri10-1868} catalogue\footnote{http://irsa.ipac.caltech.edu/Missions/wise.html} contains fluxes at 3.4, 4.6, 12, 22~$\mu$m for several cluster members (5-$\sigma$ sensitivity of 0.25, 0.35, 3.0, 18.0~mJy), which is particular useful outside \textit{Spitzer} coverage. Source blending within the 12~arcsec beam of \textit{WISE} at 22~$\mu$m is a major problem, so fluxes in this band are disregarded for SED fitting.

Wide-field near-IR maps ($28\times28$~arcmin) were obtained using CTIO/NEWFIRM (PI: Rawle; 2011A-3095), with a mean seeing of 1.7~arcsec. Total integration times of 3000~s and 6500~s were achieved in $J$- and $K_{\rm s}$-bands, yielding a 5-$\sigma$ sensitivity of $\sim$7~$\mu$Jy ($m_{\rm AB}$ $=$ 21.0, 20.6 mag respectively). Science images were downloaded from the CFHT archive,\footnote{http://www3.cadc-ccda.hia-iha.nrc-cnrc.gc.ca/cfht/} with photometry calibrated to 2MASS.

\subsection{Optical, ultraviolet and X-ray}

Galaxy morphologies (Section \ref{sec:morphology}) are based on three-band (435, 606, 814~nm) imaging from \textit{HST}/ACS WFC1, obtained via the HFF webpage. Images have 30~mas~pixel$^{-1}$, covering a $4.8\times3.5$~arcmin central region (PI: Dupke, ID 11689; HFF PI: Lotz, 13495) and a $3.5\times3.5$~arcmin parallel field (HFF PIs: Siana, Lotz, IDs 13389, 13495). Further HFF data (scheduled for May--July 2014) will increase depth, but not significantly add to spatial coverage.

$U$-band (365~nm) imaging is from the ESO2.2m/WFI, with raw data downloaded from the ESO archive.\footnote{http://archive.eso.org} Re-reduction used an updated version of \textit{Alambic} \citep{van02-123}. The average seeing was $\sim$1.5~arcsec FWHM and the final image has a 5-$\sigma$ sensitivity m$_{\rm AB}$$\sim$21.5 mag.

UV observation originates from the \textit{Galaxy Evolution Explorer} (\textit{GALEX}) All Sky Imaging Survey (AIS).\footnote{http://galex.stsci.edu} The SPIRE footprint is located within the overlap of two 500~s tiles, yielding a near-UV band ($NUV$; 2267~nm) magnitude limit of $m_{\rm AB}$ $=$ 22.5~mag. The PSF FWHM is 5~arcsec.

The \textit{Chandra} X-ray data (0.5--7 keV; $L_X \ga 10^{40} $ erg~s$^{-1}$ at $z\sim0.3$) were presented in \citet{owe11-27}. The map is reproduced in Figure \ref{fig:field}.

\subsection{Spectroscopic redshifts}

Spectroscopic redshifts are primarily from VLT/VIMOS \citep{bra09-947} and AAT/AAOmega \citep{owe11-27}, with a few additional objects observed using NTT/EMMI \citep{bos06-461} and Magellan/LDSS2 (PI: Egami). The final catalogue comprises 1183 individual sources within 14~arcmin of the cluster centre, including 447 cluster member galaxies $0.273 \le z \le 0.339$ ($cz_{\rm c}\pm10000$ km s$^{-1}$; following \citealt{owe11-27}). There are 194 cluster members within the central 4~arcmin (1.1~Mpc).

The 298 foreground galaxies with spectroscopic redshifts are used as a field sample in Section \ref{sec:tdust}, and are analysed identically to the cluster sample. Background galaxies are excluded due to increased uncertainties from the lensing effect.

\section{Analysis}
\label{sec:analysis}

\subsection{Multi-band catalogues}
\label{sec:cats}

Band-merged catalogues were produced within the Rainbow Cosmological Surveys Database framework.\footnote{https://rainbowx.fis.ucm.es} Source-matching between spectroscopic, UV, optical and near-IR (NEWFIRM, \textit{WISE}, IRAC) catalogues follow a trivial nearest-counterpart algorithm. The deep IRAC data allows excellent astrometric co-alignment of the MIPS image via bright isolated 24~$\mu$m sources. Despite the 6~arcsec FWHM PSF, every MIPS source is robustly associated with a single counterpart at shorter wavelengths. Increased source blending in \textit{Herschel} bands calls for the use of a simultaneous PSF-fitting photometry technique. We use the master \textit{Spitzer} catalogue as priors where available, and revert to direct detection in SPIRE beyond the \textit{Spitzer} coverage.

Integrated cluster properties (Section \ref{sec:cluster}) are taken from a central 4~arcmin-radius (1.1~Mpc) region, where UV, IR and spectroscopic coverage is most complete. The area encompasses 73 sources detected in at least two \textit{Herschel} bands, of which 37 have a spectroscopic redshift, with 20 confirmed cluster members. The deep 24~$\mu$m imaging probes marginally fainter dust than \textit{Herschel}, so three MIPS-detected (but \textit{Herschel}-undetected) cluster members are also included. There are 46 $NUV$-detected sources, of which 42 have a spectroscopic redshift, with 14 galaxies in the cluster. In all, this central region contains 27 star-forming cluster galaxies (i.e. detected in UV and/or IR).

At larger radii, IR-bright cluster members are restricted to directly-detected SPIRE sources matched to the optical via NEWFIRM and \textit{WISE} data only. We are very conservative in confirming counterparts without full \textit{Spitzer} and PACS coverage, which significantly impacts completeness. Seven secure \textit{Herschel}-detected spectroscopic cluster members are identified, two of which are aided by IRAC coverage. We also locate one additional MIPS/IRAC-detected cluster member without SPIRE flux. Spectroscopy confirms that 24 \textit{GALEX} sources beyond the 4~arcmin radius are cluster members. In total, the outer region has 26 confirmed star-forming cluster members.

\subsection{IR SEDs and SFR$_{\rm IR}$}
\label{sec:irsed}

The characteristic dust temperature ($T_{\rm dust}$) of the IR component is calculated via the best-fitting single-temperature modified blackbody. Dust temperature is degenerate with the emissivity index $\beta$ \citep[e.g.][]{bla03-733}, so we assume $\beta=1.5$. Using $\beta=2.0$ instead would systematically decrease $T_{\rm dust}$ by $\sim$10\%.

Total IR luminosity ($L_{\rm IR}$) is calculated by integrating the best-fitting \citet{rie09-556} template (allowing an overall normalisation) within the rest-frame wavelength range $\lambda=5-1000$~$\mu$m. Obscured SFR (SFR$_{\rm IR}$) follows directly via the \citet{ken98-189} relation, modified to match a \citet{kro02-82}-like initial mass function as in \citet{rie09-556}.\footnote{SFR$_{\rm IR}$ [$M_{\sun}$ yr$^{-1}$] $=$ 0.66$\times$SFR$_{\rm IR,K98}$ $=$ 1.14$\times$10$^{-10}$ $L_{\rm IR}$ [L$_{\sun}$]} In regions with PACS data, the SFR$_{\rm IR}$ limit at the cluster redshift is SFR$_{\rm IR}$$\sim$2~$M_{\sun}$~yr$^{-1}$. With SPIRE only, this increases to $\sim$4~$M_{\sun}$~yr$^{-1}$.

\textit{Chandra} imaging reveals one X-ray point source associated with an IR-bright cluster galaxy (HLS001427--302344). The X-ray luminosity ($L_X$ $\sim$ $10^{41}$ erg~s$^{-1}$) suggests the presence of an active galactic nucleus (AGN), and optical spectroscopy confirms a low luminosity Seyfert 1 \citep{owe12-23}. Hence for this galaxy, we find the best-fitting sum of a \citeauthor{rie09-556} template and the mean low-luminosity AGN from \citet{mul11-1082}. The AGN dominates 24~$\mu$m flux ($\sim$75\%), but has negligible influence in the \textit{Herschel} bands.

For sources detected by MIPS but not \textit{Herschel}, we extrapolate SFR$_{\rm IR}$ directly from the 24~$\mu$m flux using the \citeauthor{rie09-556} formula at $z=0.308$.\footnote{SFR$_{\rm IR}$ [$M_{\sun}$ yr$^{-1}$] $=$ 7.8$\times$10$^{-10}$ $L_{24}$ [$L_{\sun}$]} $T_{\rm dust}$ remains unconstrained.

All uncertainties are calculated by performing 1000 Monte Carlo simulations based on the estimated flux errors.

\subsection{SFR$_{\rm UV}$}
\label{sec:sfruv}

The unobscured SFR (SFR$_{\rm UV}$) is derived directly from the UV luminosity following \citet{dad04-746}.\footnote{SFR$_{\rm UV}$ [$M_{\sun}$ yr$^{-1}$] $=$ $L_{150\mu m}$ [erg s$^{-1}$ Hz$^{-1}$] / 8.85$\times$10$^{27}$} For $z=0.3$ cluster members we use the \textit{GALEX} $NUV$ band to approximate rest frame 150~$\mu$m. The mean $NUV$-band sensitivity within the SPIRE footprint corresponds to a SFR$_{\rm UV}$ limit of SFR$_{\rm UV}$$\sim$1~$M_{\sun}$~yr$^{-1}$.

\subsection{SFR$_{\rm UV+IR}$}
\label{sec:sfrtot}

Total SFR (SFR$_{\rm UV+IR}$) is simply the sum of the two SFRs (no extinction corrections are applied). However, for galaxies undetected in either UV or IR, the unknown SFR is estimated as follows. For UV-bright galaxies undetected at 24~$\mu$m or longer wavelengths, we derive the obscured SFR$_{\rm IR}$ from the internal extinction. The extinction is estimated via the UV slope (observed $NUV$--$U$ $\sim$ rest frame 150--230~$\mu$m), following the general trend of the Local Volume Legacy survey from \citet{dal09-517} (their Figure 13). Given the intrinsic scatter, uncertainties are $\sim$50\%. The largest estimated SFR$_{\rm IR}$ is for GLX001339--303015 ($7.3\pm3.3$~$M_{\sun}$~yr$^{-1}$), which lies outside of the southwestern extremity of the SPIRE coverage. For all galaxies, the estimated SFR$_{\rm IR}$ is consistent with the IR non-detections. 

Estimating SFR$_{\rm UV}$ for IR-bright galaxies undetected in $NUV$ (and typically also undetected in $U$), is more challenging. However, the detection limit indicates that SFR$_{\rm UV}<1$~$M_{\sun}$~yr$^{-1}$, so we simply assume that SFR$_{\rm UV+IR}$$=$SFR$_{\rm IR}+0.5$~$M_{\sun}$~yr$^{-1}$. Generally in these cases, SFR$_{\rm UV}$ constitutes $<$20\% of the total.

\subsection{Stellar mass}
\label{sec:mstel}

Galaxy stellar mass ($M_*$) for all spectroscopic cluster member galaxies is estimated from IRAC (or \textit{WISE}, outside IRAC coverage) 3.6 and 4.5~$\mu$m fluxes, following the relation of \citet{esk12-139} at $z=0.308$.\footnote{$M_*$ [$M_{\sun}$] $=$ $10^{14.65}S^{2.85}_{\rm 3.6 \mu m}S^{-1.85}_{\rm 4.5 \mu m}$, where $S_\lambda$ are in Jy} Stellar mass is unconstrained for 22/447 cluster members (six within the \textit{GALEX}-detected SF catalogue) as they are beyond IRAC imaging and are undetected by \textit{WISE}.

\section{Results}
\label{sec:results}

\subsection{Star-forming cluster galaxies}

The star-forming (spectroscopically-confirmed) cluster population comprises 53 galaxies: 16 UV+IR, 15 IR-only, 22 UV-only. These sources are listed in Table \ref{tab:sources}. The total SFR attributable to cluster members is SFR$_{\rm UV+IR}$ $=$ $343\pm10$~$M_{\sun}$~yr$^{-1}$, with $\sim$70\% observed in the IR (obscured star formation). The IR imaging reveals that there are no ULIRGs ($L_{\rm IR} > 10^{12}$ L$_{\sun}$) in A2744, and all eight LIRGs ($L_{\rm IR} > 10^{11}$ L$_{\sun}$) have $L_{\rm IR} < 2\times10^{11}$~L$_{\sun}$ (SFR$_{\rm IR}$ $<$ 20~$M_{\sun}$~yr$^{-1}$). These numbers are typical of the volume limited Local Cluster Substructure Survey (LoCuSS) at $z\sim0.15-0.3$, which reports on average 0.1~ULIRG and $\sim$5~LIRGs per cluster \citep{hai13-126}.

\subsubsection{Optical morphology of SF galaxies}
\label{sec:morphology}

\begin{figure*}
\centering
\includegraphics[width=177mm]{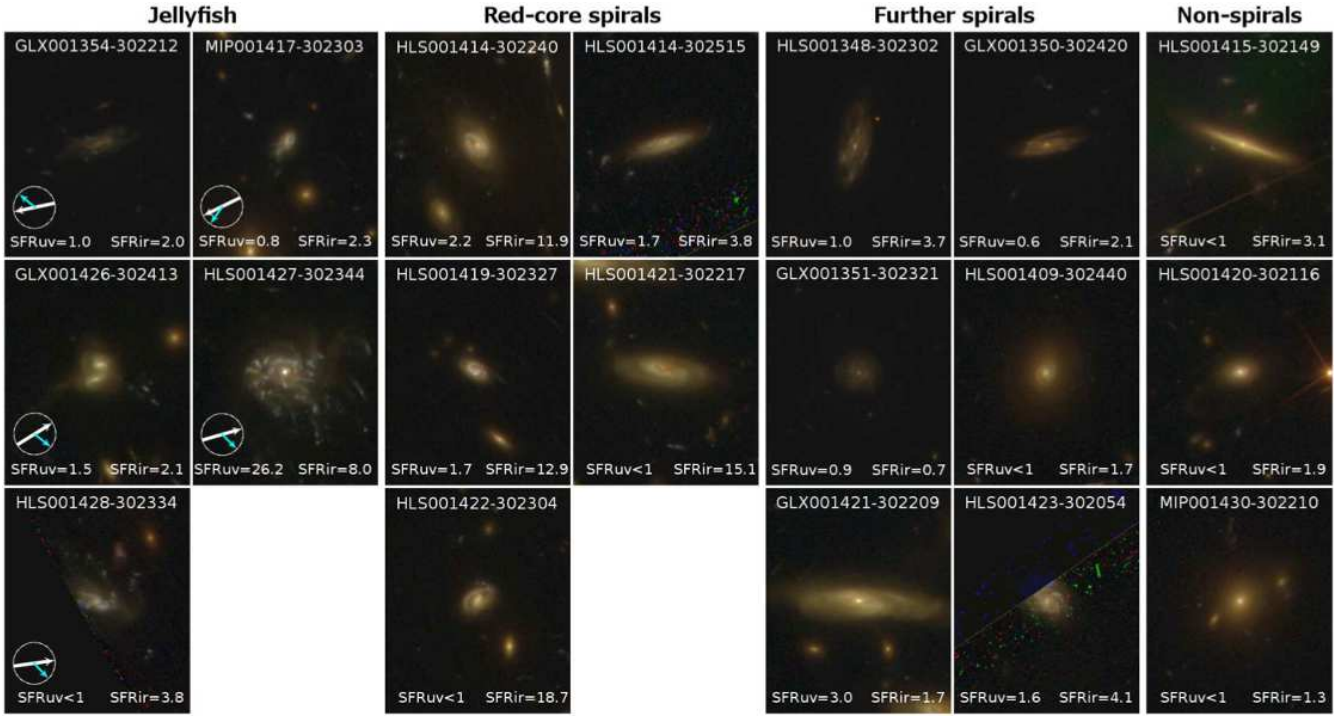}
\caption{Three-band, optical thumbnails (13$\times$16~arcsec; $\sim$59$\times$73~kpc) for star-forming cluster members covered by \textit{HST}/ACS imaging, with SFRs in $M_{\sun}$~yr$^{-1}$. The galaxies are grouped by morphology. For jellyfish, thick white arrows point towards the nominal cluster centre and thinner cyan arrows indicate the approximate orientation of the primary stripped tail (see Section \ref{sec:substructure}).}
\label{fig:thumbs}
\end{figure*}

High resolution optical data can help differentiate evolutionary mechanisms. For example, close companions and looping trails indicate galaxy--galaxy interaction, while isolated disturbance suggest a role for the cluster potential. \textit{HST}/ACS observations, including recently obtained \textit{HST Frontier Fields} data, cover nineteen star-forming cluster galaxies (Figure \ref{fig:thumbs}). We classify the morphologies of these objects to aid discussion in later sections.

Several galaxies exhibit multiple star-forming trails, blue knots and asymmetric morphologies consistent with `jellyfish' galaxies in the local Universe (first class, Figure \ref{fig:thumbs}). Three (MIP001417--302303, GLX001426--302413, HLS001427--302344) were previously reported by \citet{owe12-23} (their F1228, F0237, F0083 respectively). A fourth, HLS001428--302334, is located very close to the \citeauthor{owe12-23} jellyfish with a similar optical colour and disturbed morphology, but was previously ignored due to the ACS image edge.  Finally, GLX001354--302212 is a newly identified jellyfish (with a $>$20~kpc blue tail) to the west of the cluster in the HFF parallel field. HLS001427--302344 is the most spectacular jellyfish, with a large number of blue tentacles, the highest total SFR (SFR$_{\rm UV+IR}=34.2\pm1.3$~$M_{\sun}$~yr$^{-1}$) in the cluster and an unusually large ratio of unobscured-to-obscured star formation (SFR$_{\rm UV}$/SFR$_{\rm IR}$$\sim$3.3) for such an active galaxy. Only one other IR-detected galaxy (HLS001430--302433) exhibits SFR$_{\rm UV}$$>$SFR$_{\rm IR}$ (SFR$_{\rm UV}$/SFR$_{\rm IR}$$\sim$1.8), located $<$0.25~Mpc (in the plane of the sky) from the group of three jellyfish, but unfortunately beyond \textit{HST} coverage so the morphology is unknown. Four of five confirmed jellyfish are located together near the cluster centre, where cluster in-fall is unlikely to be the driving mechanism. We discuss the location of these galaxies, and the orientation of their blue tails, in Section \ref{sec:substructure}.

Spiral is the predominant morphology for star forming cluster members (11/19), exhibiting a wide range of arm/disc contrast and arm tightness. By eye, we identify a distinct morphological sub-set of spirals with optically-red cores, broad dusty (not blue) arms and no sign of recent galaxy interaction (second class, Figure \ref{fig:thumbs}). This group of five galaxies includes $\sim$50\% of the star formation (and all of the LIRGs) covered by ACS imaging and have a mean SFR$_{\rm UV}$/SFR$_{\rm IR}$$\sim$0.16. They are likely to represent the as-yet unstripped in-falling spiral population. The other six spirals exhibit lower SFRs and a higher mean SFR$_{\rm UV}$/SFR$_{\rm IR}$$\sim$0.72.

The remaining sources have non-spiral morphologies and low total SFRs ($\la 4$~$M_{\sun}$~yr$^{-1}$). HLS001420--302116 has a long, faint tidal loop (extending $\sim$25 kpc to the southeast) suggestive of a past galaxy encounter rather than disturbance driven by the cluster potential. HLS001415-302149 is an edge-on disc galaxy, while MIPS001430-302210 appears to be a smooth, compact early-type.

\subsubsection{Specific SFR (sSFR)}

SFR per unit stellar mass (specific SFR; sSFR), quantifies the efficiency of star formation, which generally increases with redshift \citep[e.g.][]{per05-82}. At fixed redshift, sSFR decreases with increasing stellar mass (i.e. less massive galaxies are relatively more efficient; \citealt{per08-234}), providing a main sequence (MS) of galaxies forming stars via the `normal' spatially extended mode.

\begin{figure}
\centering
\includegraphics[viewport=20mm 0mm 142mm 143mm,height=84mm,angle=270,clip]{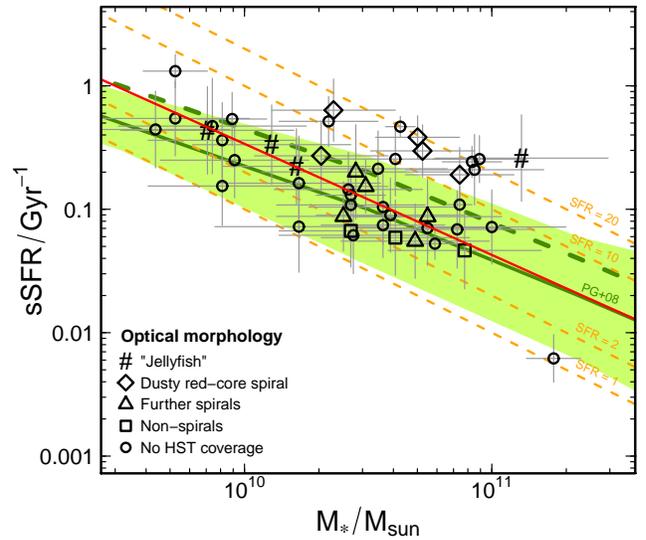}
\caption{Specific total (UV+IR) star formation rate (sSFR) versus stellar mass ($M_*$) for cluster galaxies. The solid red line shows the best-fitting main sequence for A2744 (orange dashed lines trace constant SFRs in $M_{\sun}$~yr$^{-1}$). The solid dark green line represents the typical trend at $z=0.3$ (\citealt{per08-234}, Figure 8), with 1$\,\sigma$ uncertainty shown by green shading. The steeper relation in A2744 is caused by the SFR limit becoming increasingly restrictive at the lowest masses. Galaxies with at least twice the sSFR of the fiducial main sequence (those above the dashed green line) can be considered as starbursts \citep{elb11-119}. Symbols indicate optical morphology, where HST imaging is available.}
\label{fig:ssfr}
\end{figure}

We derive sSFR from SFR$_{\rm UV+IR}$/$M_*$, and hence uncertainties in sSFR and $M_*$ are highly anti-correlated along the lines of constant SFR. The majority of SF cluster galaxies in A2744 are located on the MS (Figure \ref{fig:ssfr}). The best-fitting linear relation to the cluster sources yields a steeper relation than \citeauthor{per08-234}, but incompleteness below SFR$\la$2~$M_{\sun}$~yr$^{-1}$ is increasingly important at lower $M_*$. Here, we primarily employ Figure \ref{fig:ssfr} to differentiate modes of star formation. Starbursts, a highly efficient mode of concentrated star formation, are often defined as those with sSFRs at least twice that of the intrinsic mass-dependent MS \citep[e.g.][]{elb11-119}. The jellyfish galaxy HLS001427--302344 is starbursting, presumably observed near peak efficiency as SFR$_{\rm UV}$/SFR$_{\rm IR}$ suggests that the dusty gas reservoir has been almost completely stripped and exhausted. HLS001430--302433, the only other IR-bright galaxy with SFR$_{\rm UV}$/SFR$_{\rm IR}>1$ (Section \ref{sec:morphology}), also has a starburst-like sSFR.

The dusty, red-core spirals are also located in the starburst region of Figure \ref{fig:ssfr}, exhibiting a mean sSFR four times that of the MS ($\langle$sSFR/sSFR$_{\rm MS}$$\rangle$ $=$ 4.2 $\pm$ 0.6). The remaining confirmed (`further') spirals are on the main sequence ($\langle$sSFR/sSFR$_{\rm MS}$$\rangle$ $=$ 1.0 $\pm$ 0.3), although we note that sSFR could be underestimated if the SFR of localised starbursts is counter-balanced by large-scale quenching throughout a galaxy disc \citep[e.g.][]{mer13-1747}.

\subsubsection{Characteristic dust temperature ($T_{\rm dust}$)}
\label{sec:tdust}

\begin{figure}
\centering
\includegraphics[viewport=10mm 0mm 117mm 146mm,height=84mm,angle=270,clip]{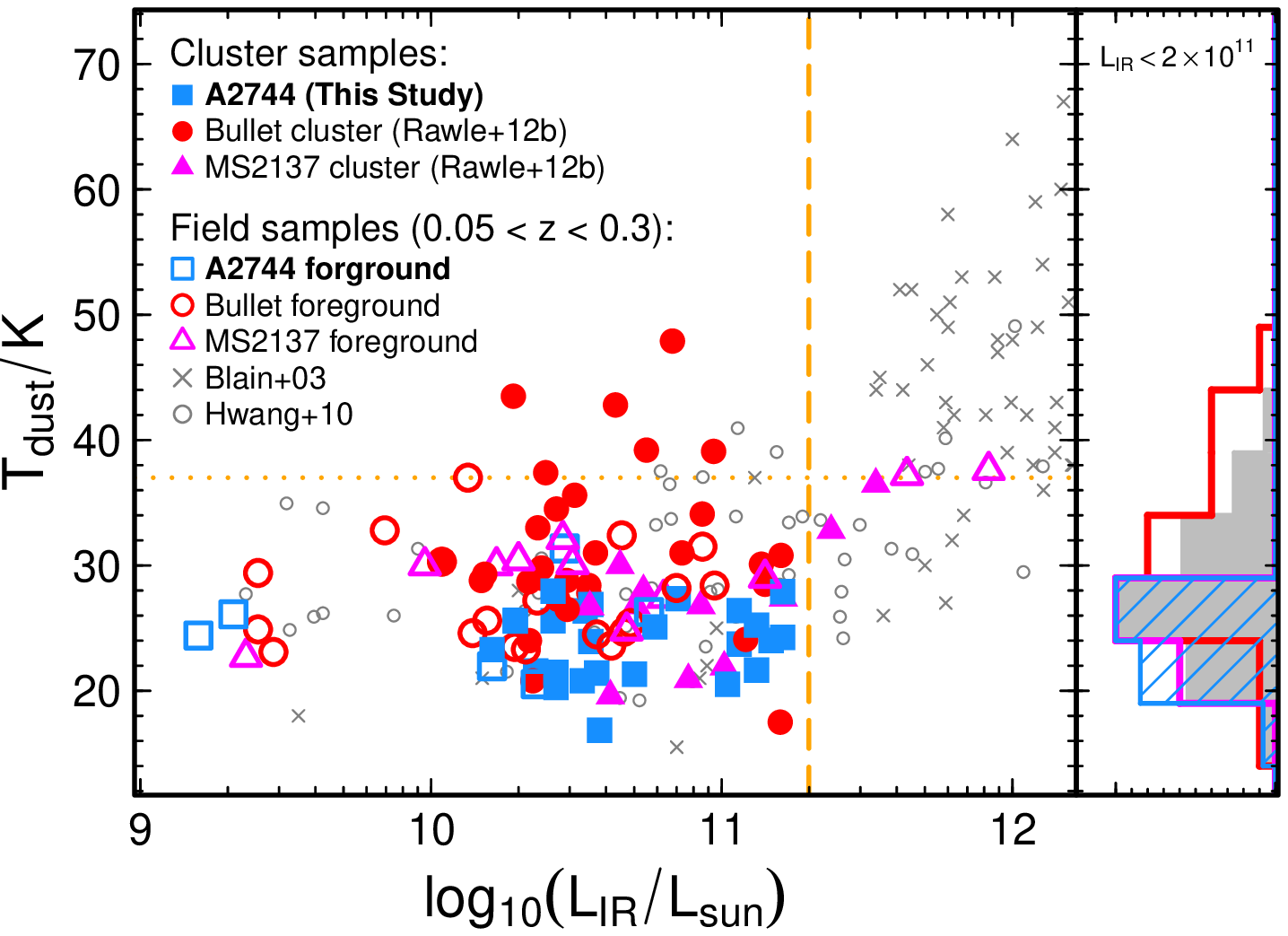}
\caption{\textit{Main panel:} Characteristic dust temperature ($T_{\rm dust}$) versus IR luminosity ($L_{\rm IR}$) for galaxies in A2744 (blue filled squares). Cluster comparisons are from \citet{raw12-106}: Bullet cluster merger (red filled circles) and the relaxed MS2137 (magenta filled triangles). Field galaxies are from cluster foregrounds (corresponding open symbols), \citet{bla03-733} and \citet{hwa10-75} (grey symbols). \textit{Right panel:} $T_{\rm dust}$ distribution for $L_{\rm IR} < 2\times10^{11} L_{\sun}$: individual clusters (coloured open histograms) and {\em combined} foreground (filled grey). The unusually warm sources are unique to the Bullet cluster.}
\label{fig:lfir_tbb}
\end{figure}

In the merging Bullet cluster ($z=0.296$), \citet{raw12-106} measured high characteristic dust temperatures for several sub-LIRGs, interpreted as a post-stripped population with cooler, outer gas/dust removed. A lack of similar sources in low density environments or a relaxed control cluster (MS2137, $z=0.313$) indicated a stripping mechanism driven by the cluster merger. A2744 offers an opportunity to investigate further. The unusually warm Bullet cluster sources are immediately obvious in the $L_{\rm IR}$--$T_{\rm dust}$ diagram (Figure \ref{fig:lfir_tbb}), which exhibits a very shallow correlation below the sharp upturn at $L_{\rm IR} \sim 2\times10^{11} L_{\sun}$. A2744 contains no similar warm galaxies, with the highest measured temperature $<$28~K. While the $T_{\rm dust}$ distribution for the low IR luminosity sample in the Bullet cluster (\{mean,rms\} $=$ 31.4, 6.6~K) exhibits a tail to warm temperatures, the peaky distribution of A2744 closely resembles that of MS2137 (\{mean,rms\} $=$ 23.8, 2.8~K; 24.8, 3.1~K respectively).

The $T_{\rm dust}$ measurements for galaxies in A2744 and the Bullet cluster follow identical procedures with the same modified blackbody parameterisation ($\beta=1.5$). The mean uncertainty on the temperature is 2.5~K in A2744 and 2.0~K for the Bullet cluster. Physically, the similarity between the cluster mergers also offers no obvious explanation for the population difference. Both are viewed just after core passage ($\sim$0.1--0.3 Gyr; \citealt{chu10-1536,mer11-333}) and the shock front in each travels through the halo ICM at approximately Mach 3 \citep{mar06-723,owe11-27}. A2744 is $\sim$20\% less massive, and has a larger X-ray cooling time and higher central entropy, indicating a more disruptive merger \citep{hud10-37}. The unusual sources in the Bullet cluster reside at local densities which are not atypical for A2744. Further examples of sources with unusual $T_{\rm dust}$ are required to constrain their origin.

\subsection{Cluster substructure and star formation}
\label{sec:substructure}

\begin{figure}
\centering
\includegraphics[viewport=2mm 0mm 148mm 145mm,height=84mm,angle=270,clip]{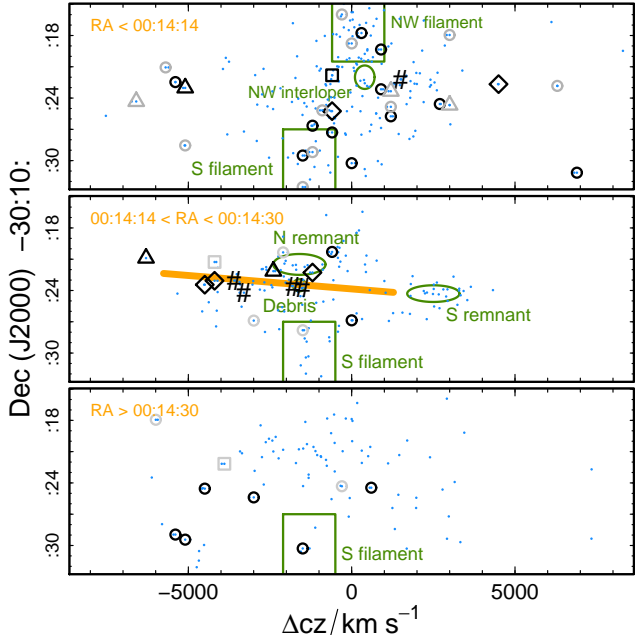}
\caption{Declination versus redshift ($\Delta cz$ with respect to $z=0.308$) in three RA slices. The central slice spans 4~arcmin around the nominal cluster RA, thereby containing the primary north--south merger. Star-forming galaxies are represented by symbols, as in Figure \ref{fig:ssfr} (black have SFR$_{\rm UV+IR}$ $>$ 3~$M_{\sun}$~yr$^{-1}$; otherwise grey). \textit{All} cluster galaxies with a spectroscopic redshift are shown by blue dots. Cluster substructure is marked via green regions analogous to Figure \ref{fig:field}. The estimated dynamic axis of the southern remnant is marked in orange.}
\label{fig:cones}
\end{figure}

We now consider the location of cluster star formation (Figures \ref{fig:field} and \ref{fig:cones}). \citet{owe11-27} used a Kaye's Mixture Model algorithm to assign galaxies to sub-components. Although 25\% were allotted to the two central remnant cores, only one of those is detected in the UV or IR: HLS001420--302116 (northern remnant). \textit{HST} imaging shows that the galaxy has a faint tidal loop, suggesting harassment, but no obvious near neighbour or dual-nucleus, arguing against a recent interaction. Galaxies within the remnant cores appear wholly undisturbed by the cluster merger, and most SF cluster members are located in the post-merger halo.

The two remnant cores exhibit a mutual $\Delta z=0.014$, but redshift is not a true spatial dimension, combining distance from the observer \textit{and} peculiar velocity. The southern remnant core (the `bullet') is travelling away from the observer ($v_{\rm los}\sim2400$~km~s$^{-1}$; \citealt{owe11-27}) and is closer to the northern remnant core than Figure \ref{fig:cones} may suggest. A trail of SF galaxies (including four jellyfish and three LIRG-type red-core spirals) traces the expected past trajectory of the bullet, clearly visible in the central panel of Figure \ref{fig:cones}. The galaxies are unlikely to be debris stripped from the supersonic southern remnant itself, as they are mostly blue-shifted with respect to the northern remnant. Rather, we suggest that many of these galaxies were located within the halo of the pre-merger northern cluster, and were encountered by the shock front associated with the passage of the bullet. 

Morphological characteristics such as the jellyfish blue tails originate from the stripping of material by extreme ram pressure. The potency of the mechanism will be a function of intrinsic galaxy properties (mass, SFR) and disc/shock orientation. Three of the jellyfish within the trail highlighted above (GLX001426--302413, HLS001427--302344, HLS001428--302334) exhibit blue tails pointing in the direction of the dynamic axis of the cluster merger (Figure \ref{fig:thumbs}). This orientation is perpendicular to the radial line of the cluster and therefore inconsistent with an in-fall scenario, in which tails would tend to point away from the cluster centre \citep[e.g.][]{smi10-1417}. Furthermore, the merger core-passage phase occurred 120--150~Myr ago \citep{mer11-333}, which corresponds well with the stellar population age of the young stars in the jellyfish ($\sim$100~Myr ago; \citealt{owe11-27}). We conclude that at least these three galaxies, located very close to the past trajectory of the bullet-like remnant core, have been stripped by the passing shock front.

The blue tail of the fourth central jellyfish (MIP001417--302303) is orientated in a different direction. Pointing towards the cluster centre, it is inconsistent both with the shock front motion and with stripping during in-fall first pass. Noting the proximity of the nominal cluster centre ($<$150~kpc) we suggest that the galaxy is observed moving outwards soon after the first pass. The final jellyfish (GLX001354--302212) is located away from the primary merger, surrounded by several other SF galaxies. The proximity of the filament suggests that these galaxies may be a small in-falling group, with the jellyfish stripped by an increasingly dense ICM. However, the blue tail points perpendicular to the radial line, towards the northeast (Figure \ref{fig:thumbs}). This orientation is consistent with the proposed northeastern motion of the northwestern interloper \citep{owe11-27}, raising the intriguing alternative scenario that the clump of star forming galaxies may be associated with the passage of this third sub-component.

At larger radii ($>$4~arcmin), $\sim$70\% of the cluster galaxies (by number or stellar mass) reside within the southwestern half of the cluster. This is not a consequence of spectroscopic coverage, which is symmetric about the cluster core, but rather the primary evidence for the existence of large-scale filaments interfacing with the cluster to the south and northwest \citep{bra09-947}. This southwestern half contains $\sim$95\% of known cluster SFR beyond 4~arcmin, and the enhancement is not due to IR/UV coverage (Figure \ref{fig:field}). The filaments themselves account for $\sim$65\% of the SFR (from 40\% of the stellar mass) at large radii, including 3 LIRGs (SFR$_{\rm IR}$ $=$ 12.9, 15.0, 16.7~$M_{\sun}$~yr$^{-1}$). This overabundance of activity is consistent with preferential location of peripheral star formation in filaments \citep[e.g.][]{fad08-9}.

Generally, the distribution of the SF population appears typical of cluster environments. However, the location of at least three extreme jellyfish galaxies suggests an evolutionary path connected with the passage of cluster merger induced shock fronts. In the next section we examine whether these transformations significantly affect the global star formation characteristics of cluster.

\subsection{Total cluster star formation}
\label{sec:cluster}

We quantify the overall effect of the A2744 merger on the galaxy population by comparing to other clusters at similar redshift. For this purpose, we compute the mass-normalised total SFR$_{\rm IR}$ ($\Sigma_{\rm SFR}$) and fraction of SF cluster galaxies ($f_{\rm SF}$) in each cluster. We include only the central 1.1~Mpc (4~arcmin in A2744), isolating the centrally-located effects of any recent merger and reducing the influence of surrounding in-fall regions and large-scale structure.\footnote{Mass-dependent scale lengths, such as the virial radius or $r_{500}$, were considered but prove hard to define for unrelaxed cluster mergers. We instead choose to impose a fixed radial limit for all clusters. For comparison, LoCuSS clusters have $r_{500}$$=$0.8--1.6~Mpc \citep{hai13-126}.} The chosen central region of A2744 also benefits from homogeneous UV--IR coverage. Previous studies rarely include unobscured SF, so for a fair comparison we are restricted to IR-detected SF galaxies and SFR$_{\rm IR}$, although SFR$_{\rm UV}$ is discussed where appropriate.

We first explore whether the cluster merger increases the fraction of galaxies with obscured star formation in the total cluster population ($f_{\rm SF}$). We follow \citet{hai13-126}, who consider only galaxies brighter than $M_K < {K^*}+1.5$~mag.\footnote{Corresponding to --23.1~mag (Vega) or $S>45$~$\mu$Jy for A2744, well within our NEWFIRM $K$-band sensitivity} Within $R<4$~arcmin of A2744 there are 428 $K$-band sources brighter than this limit. Spectroscopy confirms that 187 are cluster members, while definitively placing 42 outside. Of the remaining 199 sources, 38 have photometric redshifts \citep{bus02-787} inconsistent with the cluster ($z<0.2$ or $z>0.4$), so the cluster population will number between 187 and 348 ($=$428--42--38). Adopting the \citeauthor{hai13-126} definition of star forming (SFR$_{\rm IR}$ $\ge$ 3), A2744 contains 15 SF galaxies, yielding $f_{\rm SF} = 0.080^{+0.010}_{-0.037}$.

\begin{figure}
\centering
\includegraphics[viewport=20mm 0mm 142mm 143mm,height=84mm,angle=270,clip]{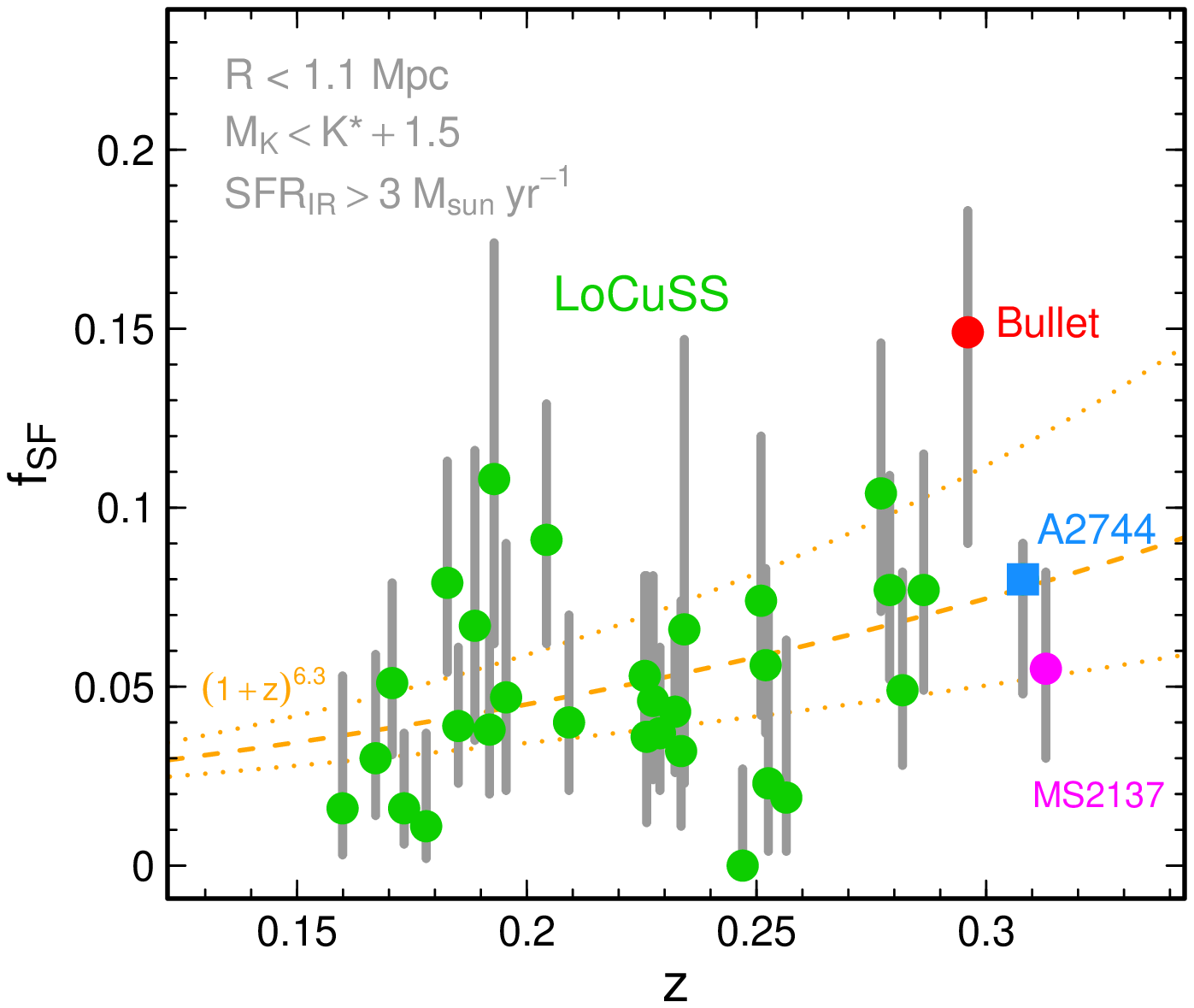}
\caption{Fraction of SF galaxies within $R<1.1$~Mpc ($f_{\rm SF}$). A2744 (filled blue square) is compared with the Bullet merger (red) and MS2137 (magenta) from \citet{raw12-106}, and the LoCuSS sample (green; \citealt{hai13-126}), all recalculated to match our radial limit. The orange lines show the best-fitting evolutionary trend: $f_{\rm SF} \propto (1+z)^n$, $n=6.3^{+1.7}_{-1.5}$ \citep{hai13-126}. In A2744, $f_{\rm SF}$ matches the general cluster population, but for the Bullet, the fraction is enhanced.}
\label{fig:fsf}
\end{figure}

Figure \ref{fig:fsf} shows that A2744 lies on the best-fitting evolutionary trend from the LoCuSS sample, which includes both relaxed and recently disturbed clusters \citep{san09-1698,raw12-29}, alongside the relaxed MS2137. The merging Bullet cluster exhibits a very weak ($\sim$1$\,\sigma$) enhancement in $f_{\rm SF}$ \citep{raw12-106}. The possibility of a general correlation between cluster morphology and global star formation properties will be explored for the LoCuSS sample by Scott et~al. (in preparation).

\begin{figure}
\centering
\includegraphics[viewport=20mm 0mm 142mm 143mm,height=84mm,angle=270,clip]{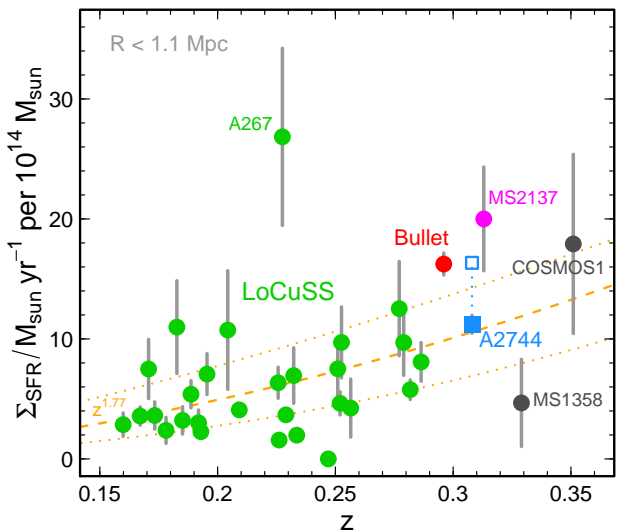}
\caption{Mass-normalised total SFR$_{\rm IR}$ ($\Sigma_{\rm SFR}$) within $R<1.1$~Mpc, versus cluster redshift ($z$). Symbols as in Figure \ref{fig:fsf}, with the addition of MS1358 and COSMOS--CL1 (grey; \citealt{pop12-58}). Error bars represent all uncertainties including measurement error and spectroscopic/IR incompleteness. Orange lines show the mean evolutionary trend from \citet{pop12-58}: $\Sigma_{\rm SFR} \propto z^{1.77\pm0.36}$. For A2744, the effect of including SFR$_{\rm UV+IR}$ in $\Sigma_{\rm SFR}$ is indicated by the open blue square.}
\label{fig:sig_sfr}
\end{figure}

We now progress to the mass-normalised total SFR$_{\rm IR}$ ($\Sigma_{\rm SFR}$). Within a 1.1~Mpc radius, there are 23 IR-detected spectroscopic cluster members in A2744, with a combined SFR$_{\rm IR}$ $=$ $132\pm7$~$M_{\sun}$~yr$^{-1}$. This underestimates the total cluster obscured SFR$_{\rm IR}$ due to two limits, (1) spectroscopic incompleteness and (2) the IR sensitivity. We account for (1) by examining the 36 \textit{Herschel} sources without spectroscopic redshifts. Ten have optical photometric redshifts from \citet{bus02-787}, but only two are at $z_{\rm opt} \sim 0.3$ (the others are $z_{\rm opt} > 0.6$). For the remaining 26 sources, we estimate a very approximate ``IR photo-$z$'' by fitting a typical greybody ($T_{\rm dust}$ $=$ 30 K; $\beta=1.5$) to the SED. Only one exhibits $0.1 < z_{\rm IR} < 0.6$. These three cluster candidates would have SFR$_{\rm IR}=1.8\pm0.4$, $1.8\pm0.3$, $2.4\pm0.3$~$M_{\sun}$~yr$^{-1}$. Further cluster members with SFR$_{\rm IR}$ $\ga$ 2~$M_{\sun}$~yr$^{-1}$ are unlikely.

Limit (2) excludes low-level star formation in e.g. the early-type population \citep{raw08-2097}. We estimate their total SFR by stacking 24~$\mu$m flux for 142 spectroscopic cluster members covered by MIPS data but without a formal detection. The stack has a total flux $S_{24}=3.4\pm0.1$~$\mu$Jy (SFR$_{\rm IR}$ $=$ 0.07~$M_{\sun}$~yr$^{-1}$), which represents a negligible contribution. Finally, neither test above accounts for a population faint in IR \textit{and} optical bands (i.e. excluded from spectroscopy), such as dwarf galaxies. However, it is unlikely that such galaxies would contribute significantly to the integrated cluster obscured SFR as environmental quenching is highly efficient for dwarves \citep[e.g.][]{smi12-3167}. Hence, within $R<1.1$~Mpc, the total SFR$_{\rm IR}=138\pm8$~$M_{\sun}$~yr$^{-1}$. The five LIRGs within this central region contribute 55\% of the SFR$_{\rm IR}$.

For completeness we briefly deviate from the obscured SFR and note that within the central region, total SFR$_{\rm UV}$ $=$ $63\pm3$~$M_{\sun}$~yr$^{-1}$. The stack of UV-undetected spectroscopic cluster members has a $m_{NUV,AB}=25.8\pm0.2$~mag (SFR$_{\rm UV}$ $=$ 0.05~$M_{\sun}$~yr$^{-1}$), confirming that the \textit{GALEX} sensitivity limit has a negligible effect. Unobscured star formation accounts for 30\% of the total SFR, SFR$_{\rm UV+IR}$ $=$ $201\pm9$~$M_{\sun}$~yr$^{-1}$.

Cluster mass is calculated from the ``Clusters As Telescopes'' (CATS\footnote{http://archive.stsci.edu/prepds/frontier/lensmodels/}; Richard et~al. in preparation) mass model, tightly constrained by 17 families of (strongly-lensed) multiple images of background galaxies. The mass model comprises 150 individual cluster galaxies as well as four larger dark matter halo components. Within $R<1.1$~Mpc, the total mass is $M=(1.23\pm0.04)\times10^{15}$~$M_{\sun}$. Hence for A2744, obscured $\Sigma_{\rm SFR} = 11.2\pm0.7$~$M_{\sun}$~yr$^{-1}$ per 10$^{14}$~$M_{\sun}$.

We compute the identical quantities (within 1.1~Mpc) for the Bullet cluster: SFR$_{\rm IR}=242\pm12$~$M_{\sun}$~yr$^{-1}$ \citep{raw12-106}, strong lens model mass $M=(1.49\pm0.04)\times10^{15}$~$M_{\sun}$ \citep{par12-arxiv}, $\Sigma_{\rm SFR} = 16.2\pm0.6$~$M_{\sun}$~yr$^{-1}$ per 10$^{14}$~$M_{\sun}$. This is 45\% higher than A2744.

We put these into a wider perspective in Figure \ref{fig:sig_sfr} by comparing to the 29 LoCuSS clusters, MS2137 (\citealt{raw12-106}, Richard et~al. in preparation) and two $z\sim0.3-0.4$ clusters from \citet{pop12-58}.\footnote{The \citeauthor{pop12-58} sample comprises nine clusters in total. All three at $z<0.3$ are covered by LoCuSS, and we adopt $\Sigma_{\rm SFR}$ from that analysis. The remaining four ($z\sim0.5-0.9$) are located beyond the redshift range shown in Figure \ref{fig:sig_sfr}.} A2744 lies on the best-fitting evolutionary trend derived by \citet{pop12-58} from nine clusters ($z\sim0.2-0.9$). Note that A267 is the lowest mass cluster in the LoCuSS sample, which increases the uncertainty in mass considerably, and may dampen the ability of the cluster to quench in-falling galaxies, resulting in the higher SFR. The Bullet cluster has enhanced obscured activity, although previous studies have concluded that individual galaxies immediately adjacent to the shock are unaffected \citep{chu10-1536,raw10-14}. Furthermore, the relaxed MS2137 and the unrelaxed MS1358 show marginally enhanced and depressed $\Sigma_{\rm SFR}$ respectively, suggesting that there is significant intrinsic scatter unrelated to the dynamical state of the cluster.

The current data indicates that for $z>0.3$ the overall (IR-detected) obscured SFR is not influenced by the occurrence of a recent merger. However, we have already shown that SFR$_{\rm IR}$ may not be a good tracer of the processes triggered by the cluster merger, as at least one jellyfish is known to possess a high unobscured SFR. Figure \ref{fig:sig_sfr} also displays the effect of including SFR$_{\rm UV}$ for A2744: increasing $\Sigma_{\rm SFR}$ by $\sim$50\%. With the current comparison data (which excludes SFR$_{\rm UV}$), it is difficult to ascertain whether A2744 has an unusually high SFR$_{\rm UV}$, or whether relaxed clusters are similar (systematically increasing  $\Sigma_{\rm SFR}$).

For A2744, the cluster merger has a negligible, or at least net zero, effect on obscured SFR. The large fraction of unobscured SFR ($\sim$$\frac{1}{3}$ total SFR) in the cluster, together with the known examples of extreme stripping triggered by the merger, suggests that the true effect on total SFR may be more significant.

\section{Conclusions}
\label{sec:conclusions}

This paper presents analysis of multi-wavelength data, including \textit{GALEX}, \textit{Spitzer} and \textit{Herschel} bands, for galaxies within the massive merging \textit{HST Frontier Field} cluster A2744 (Pandora's Cluster). We determine the overall (unobscured plus obscured) star formation properties for 53 cluster galaxies, with a total SFR, SFR$_{\rm UV+IR}$ $=$ $343\pm10$~$M_{\sun}$~yr$^{-1}$. From the IR, we find that the cluster contains no ULIRGs, and all eight LIRGs are fainter than $L_{\rm IR} < 2\times10^{11}$~L$_{\sun}$. A2744 contains no sub-LIRG galaxies with unusually warm ($\ga40$~K) dust, suggesting that the presence of such sources in the Bullet cluster \citep{raw12-106} may not be related to the cluster merger.

Within the central 1.1~Mpc (the \textit{Herschel} Lensing Survey PACS coverage), which encompasses both remnant cores from the primary merger as well as the innermost regions of the mixed halo population, the total SFR$_{\rm IR}=138\pm8$~$M_{\sun}$~yr$^{-1}$ (55\% from LIRGs) and SFR$_{\rm UV}=63\pm3$~$M_{\sun}$~yr$^{-1}$. Comparison with further clusters at similar redshift indicates that the merger has no significant effect on (mass-normalised) total cluster obscured SFR ($\Sigma_{\rm SFR} = 11.2\pm0.7$~$M_{\sun}$~yr$^{-1}$ per 10$^{14}$~$M_{\sun}$) or fraction of (IR-detected) star-forming galaxies ($f_{\rm SF} = 0.080^{+0.010}_{-0.037}$). The remnant core galaxy populations exhibit no signs of star formation or disturbance attributable to the cluster merger. However, there is a population of star-forming halo galaxies consistent with morphological transformation due to the passage of the shock front associated with the bullet-like southern sub-cluster. These include at least one extreme, unobscured starburst jellyfish galaxy (SFR$_{\rm UV+IR}=34.2\pm1.3$~$M_{\sun}$~yr$^{-1}$, SFR$_{\rm UV}$/SFR$_{\rm IR}$$\sim$3.3).

In A2744, the merger has a net-zero effect on the bulk obscured star formation properties of the cluster. Generally, total cluster IR dust properties (e.g. total SFR$_{\rm IR}$ and $T_{\rm dust}$ distribution) do not appear to be systematically correlated with the existence of a recent merger; A2744 and the Bullet cluster exhibit significant differences. With the lack of unobscured SFR analysis for relaxed clusters (or the Bullet), it is difficult to ascertain whether SFR$_{\rm UV}$ is unusually enhanced by the merger.

We have demonstrated that future cluster studies, particularly for merging systems where extreme ram pressure may play a significant role, require both UV \textit{and} IR imaging to adequately constrain the total star formation properties of member galaxies.

\section*{Acknowledgements}

T.~Rawle is supported by a European Space Agency (ESA) Research Fellowship at the European Space Astronomy Centre (ESAC), in Madrid, Spain. J.~Santos has received funding from the European Union Seventh Framework Programme (FP7/2007-2013) under grant agreement 267251 ``Astronomy Fellowships in Italy" (AstroFIt). C.~Haines was funded by CONICYT Anillo project ACT-1122. N.~Okabe was supported by the World Premier International Research Center Initiative (WPI Initiative), MEXT, Japan. The authors would particularly like to thank A.~Cava, F.~Combes and I.~Smail for useful discussion and M. Owers for providing the combined image from the \textit{Chandra X-ray Observatory}.

This work is based on observations made with the \textit{Herschel Space Observatory}, an ESA Cornerstone Mission with significant participation by NASA.

M. Rex and G. Walth were visiting astronomers for programme 2011A-3095 (PI: T. Rawle) at CTIO, NOAO, which is operated by AURA, under contract with NSF.

Also based on observations made with the NASA/ESA \textit{Hubble Space Telescope}, which is operated by the Association of Universities for Research in Astronomy, Inc., under NASA contract NAS 5-26555. These observations are associated with programme \#11689 and HFF DD programmes \#13389,13495. This publication also uses data products from the \textit{Wide-field Infrared Survey Explorer} (\textit{WISE}), which is a joint project of UCLA and JPL, Caltech, funded by NASA. We made use of a private version of the Rainbow Cosmological Surveys Database, operated by the Universidad Complutense de Madrid (UCM).

\appendix

\begin{table*}
\caption{Derived properties for the star-forming galaxies in A2744. Bold rows indicate galaxies within the central 4~arcmin (1.1~Mpc).}
\label{tab:sources}
\begin{tabular}{$l^c^c^c^c^c^c^c}
\hline
\multicolumn{1}{l}{ID$^1$} & \multicolumn{1}{c}{$z$} & \multicolumn{1}{c}{$T_{\rm dust}$$^2$} & \multicolumn{1}{c}{SFR$_{\rm IR}$$^3$} & \multicolumn{1}{c}{SFR$_{\rm UV}$$^4$} & \multicolumn{1}{c}{SFR$_{\rm UV+IR}$} & \multicolumn{1}{c}{$M_{\rm *}$$^5$} & \multicolumn{1}{c}{morph$^6$} \\
\multicolumn{1}{l}{} & \multicolumn{1}{c}{} & \multicolumn{1}{c}{K} & \multicolumn{1}{c}{$M_{\sun}$ yr$^{-1}$} & \multicolumn{1}{c}{$M_{\sun}$ yr$^{-1}$} & \multicolumn{1}{c}{$M_{\sun}$ yr$^{-1}$} & \multicolumn{1}{c}{$M_{\sun}$} & \multicolumn{1}{c}{} \\
\hline
HLS001329--301921 & 0.311 & 23.7 $\pm$ 3.7 & 12.9 $\pm$ 2.6 & 7.0 $\pm$ 0.6 & 19.9 $\pm$ 2.7 & 10.63 $\pm$ 0.06 & -- \\
GLX001339--303015 & 0.308 & -- & \textit{(7.3 $\pm$ 3.3)} & 3.1 $\pm$ 0.4 & 10.4 $\pm$ 3.4 & 10.61 $\pm$ 0.14 & -- \\
GLX001344--302450 & 0.312 & -- & \textit{(0.9 $\pm$ 0.4)} & 1.0 $\pm$ 0.4 & 1.8 $\pm$ 0.5 & -- & -- \\
HLS001348--302302 & 0.291 & 15.5 $\pm$ 5.5 & 3.7 $\pm$ 2.1 & 1.0 $\pm$ 0.4 & 4.7 $\pm$ 2.1 & 10.49 $\pm$ 0.09 & Sp \\
HLS001350--301746 & 0.309 & 21.3 $\pm$ 2.1 & 5.7 $\pm$ 2.1 & 1.7 $\pm$ 0.4 & 7.4 $\pm$ 2.1 & 10.54 $\pm$ 0.28 & -- \\
GLX001350--302420 & 0.286 & -- & \textit{(2.1 $\pm$ 1.0)} & 0.6 $\pm$ 0.2 & 2.7 $\pm$ 1.0 & 10.69 $\pm$ 0.13 & Sp \\
GLX001351--302321 & 0.312 & -- & \textit{(0.7 $\pm$ 0.3)} & 0.9 $\pm$ 0.3 & 1.6 $\pm$ 0.4 & -- & Sp \\
GLX001353--302639 & 0.304 & -- & \textit{(2.4 $\pm$ 1.1)} & 1.0 $\pm$ 0.3 & 3.5 $\pm$ 1.2 & 10.59 $\pm$ 0.14 & -- \\
GLX001354--302212 & 0.313 & -- & \textit{(2.0 $\pm$ 0.9)} & 1.0 $\pm$ 0.3 & 3.0 $\pm$ 1.0 & -- & JF \\
GLX001355--301559 & 0.307 & -- & \textit{(0.4 $\pm$ 0.2)} & 1.5 $\pm$ 0.4 & 1.9 $\pm$ 0.4 & 9.64 $\pm$ 0.30 & -- \\
GLX001356--301846 & 0.308 & -- & \textit{(2.2 $\pm$ 1.0)} & 0.6 $\pm$ 0.3 & 2.9 $\pm$ 1.1 & 9.72 $\pm$ 0.28 & -- \\
HLS001358--302434 & 0.317 & 25.1 $\pm$ 4.8 & 6.7 $\pm$ 2.6 & 1.4 $\pm$ 0.3 & 8.1 $\pm$ 2.6 & 10.87 $\pm$ 0.05 & -- \\
GLX001401--301757 & 0.318 & -- & \textit{(0.7 $\pm$ 0.3)} & 2.3 $\pm$ 0.4 & 2.9 $\pm$ 0.5 & 9.91 $\pm$ 0.27 & -- \\
GLX001404--302545 & 0.312 & -- & \textit{(2.5 $\pm$ 1.2)} & 1.4 $\pm$ 0.3 & 3.9 $\pm$ 1.2 & 10.74 $\pm$ 0.13 & -- \\
GLX001404--302833 & 0.291 & -- & \textit{(0.9 $\pm$ 0.4)} & 1.2 $\pm$ 0.3 & 2.1 $\pm$ 0.5 & -- & -- \\
GLX001404--303232 & 0.303 & -- & \textit{(0.7 $\pm$ 0.3)} & 1.0 $\pm$ 0.3 & 1.7 $\pm$ 0.4 & 10.44 $\pm$ 0.15 & -- \\
\rowstyle{\bfseries}
GLX001405--302104 & 0.289 & -- & \textit{(0.9 $\pm$ 0.4)} & 1.4 $\pm$ 0.3 & 2.3 $\pm$ 0.5 & 9.96 $\pm$ 0.18 & -- \\
\rowstyle{\bfseries}
HLS001406--302228 & 0.290 & 27.9 $\pm$ 4.0 & 3.0 $\pm$ 0.6 & $<$1 & 3.5 $\pm$ 0.7 & 9.87 $\pm$ 0.38 & -- \\
HLS001406--302718 & 0.306 & 21.6 $\pm$ 5.6 & 15.0 $\pm$ 2.3 & 2.8 $\pm$ 0.4 & 17.8 $\pm$ 2.3 & 10.93 $\pm$ 0.28 & -- \\
GLX001406--302931 & 0.303 & -- & \textit{(5.3 $\pm$ 2.4)} & 1.9 $\pm$ 0.4 & 7.2 $\pm$ 2.5 & 11.00 $\pm$ 0.11 & -- \\
\rowstyle{\bfseries}
MIP001407--302250 & 0.329 & -- & 0.7 $\pm$ 0.4 & $<$1 & 1.2 $\pm$ 0.5 & 10.22 $\pm$ 0.34 & -- \\
\rowstyle{\bfseries}
HLS001409--302309 & 0.311 & 21.5 $\pm$ 1.6 & 3.1 $\pm$ 0.3 & 1.9 $\pm$ 0.4 & 5.0 $\pm$ 0.5 & 10.86 $\pm$ 0.31 & -- \\
\rowstyle{\bfseries}
HLS001409--302440 & 0.318 & 15.9 $\pm$ 1.4 & 1.7 $\pm$ 0.4 & $<$1 & 2.2 $\pm$ 0.5 & 10.40 $\pm$ 0.27 & Sp \\
\rowstyle{\bfseries}
GLX001411--302511 & 0.305 & -- & \textit{(0.2 $\pm$ 0.1)} & 1.0 $\pm$ 0.3 & 1.3 $\pm$ 0.3 & 9.91 $\pm$ 0.27 & -- \\
HLS001411--303109 & 0.331 & 16.9 $\pm$ 5.6 & 4.3 $\pm$ 1.8 & $<$1 & 4.8 $\pm$ 1.8 & 9.95 $\pm$ 0.14 & -- \\
GLX001413--302911 & 0.304 & -- & \textit{(1.8 $\pm$ 0.9)} & 1.1 $\pm$ 0.3 & 2.9 $\pm$ 0.9 & 10.43 $\pm$ 0.09 & -- \\
\rowstyle{\bfseries}
HLS001414--302240 & 0.323 & 20.5 $\pm$ 0.5 & 11.9 $\pm$ 0.5 & 2.2 $\pm$ 0.5 & 14.1 $\pm$ 0.7 & 10.87 $\pm$ 0.22 & dusty Sp \\
\rowstyle{\bfseries}
HLS001414--302515 & 0.306 & 26.4 $\pm$ 3.8 & 3.8 $\pm$ 1.0 & 1.7 $\pm$ 0.3 & 5.5 $\pm$ 1.1 & 10.31 $\pm$ 0.27 & dusty Sp \\
\rowstyle{\bfseries}
HLS001415--302149 & 0.306 & 20.3 $\pm$ 1.7 & 3.1 $\pm$ 0.3 & $<$1 & 3.6 $\pm$ 0.4 & 10.89 $\pm$ 0.31 & ?? \\
\rowstyle{\bfseries}
MIP001417--302303 & 0.296 & -- & 2.3 $\pm$ 0.3 & 0.8 $\pm$ 0.3 & 3.1 $\pm$ 0.4 & 9.85 $\pm$ 0.35 & JF \\
\rowstyle{\bfseries}
HLS001418--302021 & 0.301 & 25.6 $\pm$ 3.6 & 2.2 $\pm$ 0.5 & $<$1 & 2.7 $\pm$ 0.6 & 10.56 $\pm$ 0.25 & -- \\
\rowstyle{\bfseries}
HLS001419--302327 & 0.293 & 26.4 $\pm$ 1.3 & 12.9 $\pm$ 1.1 & 1.7 $\pm$ 0.3 & 14.6 $\pm$ 1.1 & 10.36 $\pm$ 0.25 & dusty Sp \\
\rowstyle{\bfseries}
HLS001420--302116 & 0.294 & 23.3 $\pm$ 2.9 & 1.9 $\pm$ 0.3 & $<$1 & 2.4 $\pm$ 0.4 & 10.61 $\pm$ 0.30 & ?? \\
MIP001420--302749 & 0.303 & -- & 0.6 $\pm$ 0.4 & $<$1 & 1.1 $\pm$ 0.5 & 11.25 $\pm$ 0.11 & -- \\
\rowstyle{\bfseries}
GLX001421--302209 & 0.300 & -- & \textit{(1.7 $\pm$ 0.8)} & 3.0 $\pm$ 0.4 & 4.8 $\pm$ 0.9 & 10.74 $\pm$ 0.23 & Sp \\
\rowstyle{\bfseries}
HLS001421--302217 & 0.304 & 25.2 $\pm$ 1.1 & 15.1 $\pm$ 0.6 & $<$1 & 15.6 $\pm$ 0.7 & 10.72 $\pm$ 0.21 & dusty Sp \\
\rowstyle{\bfseries}
HLS001421--302652 & 0.308 & 21.6 $\pm$ 1.7 & 2.6 $\pm$ 0.3 & $<$1 & 3.1 $\pm$ 0.4 & 10.77 $\pm$ 0.12 & -- \\
\rowstyle{\bfseries}
HLS001422--302304 & 0.294 & 27.9 $\pm$ 1.4 & 18.7 $\pm$ 1.6 & $<$1 & 19.2 $\pm$ 1.6 & 10.70 $\pm$ 0.17 & dusty Sp \\
\rowstyle{\bfseries}
HLS001423--302054 & 0.287 & 26.9 $\pm$ 3.1 & 4.1 $\pm$ 0.4 & 1.6 $\pm$ 0.3 & 5.7 $\pm$ 0.5 & 10.45 $\pm$ 0.38 & Sp \\
\rowstyle{\bfseries}
HLS001424--302018 & 0.306 & 21.4 $\pm$ 1.5 & 4.2 $\pm$ 0.4 & 2.7 $\pm$ 0.4 & 6.9 $\pm$ 0.6 & 9.72 $\pm$ 0.13 & -- \\
\rowstyle{\bfseries}
GLX001426--302413 & 0.297 & -- & \textit{(2.1 $\pm$ 1.0)} & 1.5 $\pm$ 0.5 & 3.6 $\pm$ 1.1 & 10.21 $\pm$ 0.21 & JF \\
\rowstyle{\bfseries}
HLS001427--302344 & 0.303 & 27.4 $\pm$ 2.3 & 8.0 $\pm$ 1.0$^\dagger$ & 26.2 $\pm$ 0.9 & 34.2 $\pm$ 1.3 & 11.12 $\pm$ 0.35 & JF \\
\rowstyle{\bfseries}
HLS001428--302334 & 0.302 & 20.8 $\pm$ 1.4 & 3.8 $\pm$ 0.4 & $<$1 & 4.3 $\pm$ 0.5 & 10.11 $\pm$ 0.32 & JF \\
GLX001428--302653 & 0.298 & -- & \textit{(0.3 $\pm$ 0.2)} & 0.9 $\pm$ 0.2 & 1.2 $\pm$ 0.3 & -- & -- \\
\rowstyle{\bfseries}
MIP001430--302210 & 0.295 & -- & 1.3 $\pm$ 0.2 & $<$1 & 1.8 $\pm$ 0.4 & 10.43 $\pm$ 0.29 & ?? \\
\rowstyle{\bfseries}
HLS001430--302433 & 0.293 & 23.9 $\pm$ 1.2 & 4.1 $\pm$ 0.3 & 7.2 $\pm$ 0.6 & 11.3 $\pm$ 0.7 & 10.34 $\pm$ 0.20 & -- \\
HLS001430--303017 & 0.303 & 24.0 $\pm$ 1.2 & 16.7 $\pm$ 1.7 & 3.5 $\pm$ 0.5 & 20.2 $\pm$ 1.8 & 10.92 $\pm$ 0.12 & -- \\
\rowstyle{\bfseries}
HLS001431--302428 & 0.310 & 24.3 $\pm$ 0.6 & 18.5 $\pm$ 0.7 & 4.2 $\pm$ 0.3 & 22.7 $\pm$ 0.7 & 10.95 $\pm$ 0.19 & -- \\
\rowstyle{\bfseries}
HLS001431--302523 & 0.298 & 25.6 $\pm$ 2.4 & 3.0 $\pm$ 0.3 & $<$1 & 3.5 $\pm$ 0.4 & 10.43 $\pm$ 0.14 & -- \\
\rowstyle{\bfseries}
HLS001432--302419 & 0.307 & 25.6 $\pm$ 3.5 & 2.2 $\pm$ 1.2 & $<$1 & 2.7 $\pm$ 1.2 & 10.22 $\pm$ 0.33 & -- \\
GLX001434--301757 & 0.288 & -- & \textit{(0.1 $\pm$ 0.0)} & 0.8 $\pm$ 0.3 & 0.9 $\pm$ 0.3 & -- & -- \\
GLX001446--302857 & 0.290 & -- & \textit{(1.3 $\pm$ 0.6)} & 2.5 $\pm$ 0.4 & 3.8 $\pm$ 0.7 & 10.56 $\pm$ 0.21 & -- \\
GLX001447--302926 & 0.291 & -- & \textit{(2.9 $\pm$ 1.4)} & 0.9 $\pm$ 0.2 & 3.8 $\pm$ 1.4 & 10.42 $\pm$ 0.19 & -- \\
\hline
\\
\end{tabular}
\\
\raggedright
$^1$ ID from optical RA and Dec\\
$^2$ $T_{\rm dust}$ derived from modified blackbody\\
$^3$ SFR$_{\rm IR}$ calculated via... {\bf HLS: } integration of best-fit \citet{rie09-556} template; {\bf MIP:} extrapolation from 24~$\mu$m flux (Section \ref{sec:irsed}); {\bf GLX:} (\textit{bracketed}) estimation from the extinction-corrected SFR$_{\rm UV}$\\
$^4$ SFR$_{\rm UV}$ derived from \textit{GALEX} NUV band, and un-corrected for extinction\\
$^5$ $M_{\rm *}$ estimated from 3.6--4.5~$\mu$m region\\
$^6$ Optical morphology (Section \ref{sec:morphology}): JF$=$jellyfish; dusty Sp$=$dusty red-core spiral; Sp$=$further~spiral; ??$=$non-spiral. Remaining galaxies are beyond \textit{HST}/ACS coverage\\
$^\dagger$ SED fit by the sum of a \citet{rie09-556} template and the mean low-luminosity AGN from \citet{mul11-1082}
\end{table*}

\label{lastpage}
\end{document}